\documentclass[12pt,aps,prb,preprint]{revtex4}
\usepackage{graphicx}
\usepackage{amsmath}
\usepackage{amssymb}
\usepackage{slashed}
\usepackage{epsfig}
\usepackage{hyperref}
\usepackage{amsbsy}

\newcommand{\eps}{\epsilon}

\begin{document}

\title{Simple Analysis of IR Singularities at One Loop}
\author{Ambresh Shivaji }
\email{ambresh@iopb.res.in}
\affiliation{Institute of Physics, Sachivalaya Marg, Bhubaneswar - 751005, India}

\date{\today}

\begin{abstract}
 
In this article, we explore the structure of IR singularities of Feynman diagrams at one loop via power counting in loop momentum. 
The emphasis is on many known results which follow from this simple analysis.

\end{abstract}
\maketitle{}

\section{Introduction}

Our theoretical understanding of nature is largely based on perturbative methods. In Quantum Field Theory, one is interested in calculating transition amplitudes and therefore
cross-sections and decay rates for various scattering and decay processes. Such calculations are performed at a given order in perturbation theory. In the language of Feynman diagrams,  
lowest order contribution to the transition amplitude comes from only tree diagrams while higher order terms receive contribution from
loop (virtual) diagrams also. \footnote{For certain processes there does not exist any tree diagram and one loop diagrams give contribution at the lowest order. This is the case with four photon 
interaction in QED.} In literature, higher order contributions to a given transition amplitude are known as radiative corrections.\\

Computation of virtual diagrams for a given scattering or decay process, involve integration over undetermined loop momenta. These loop integrals are ill-defined as they may diverge for 
certain limiting values of momentum flowing in the loop. A well known case is of ultraviolet singularity which may arise as loop momentum becomes very large. With massless particles present in theory, these
integrals may develop infrared or more appropriately mass singularities, as well. Also, there are singularities which originate from specific phase space points such as physical and anomalous
thresholds. A systematic study of mass singularities and threshold singularities can be done via Landau Equations. {\cite{Landau59} \cite{Eden02} \cite{Zuber05}} Kinoshita ~\cite{Kinoshita62} describes mass singularities of Feynman amplitudes as pathological solutions
of Landau Equations.\\

In this article, we limit ourselves to the IR singularity of one loop diagrams, away from any threshold. Also we do not consider any exceptional phase space point, corresponding to vanishing
of any partial sum of external momenta. According to Kinoshita  ~\cite{Kinoshita62}, these are pure mass singularities, valid for all possible kinematic invariants made out of external momenta. These mass singularities can be of soft or collinear type, often described in literature. 
Under certain circumstances, which we will discuss below, the overlapping of soft and collinear singularities may also take place. 
In the following, we wish to present a pedagogical introduction of occurrence and structure of IR divergence
in one loop diagrams. In Ref.~\onlinecite{Kinoshita62} this issue is discussed thoroughly, using parametric form of loop integrals. We have treated the issue at the very elementary level and 
have show all the features of mass singularities known at one loop. \\

The most general one loop integral is of tensor type, 
in which loop momentum appears in the numerator. One loop diagrams having fermions in the loop are common examples of such integrals.
Since any tensor integral at one loop can be expressed in terms of scalar ones ($i.e.$ one loop integrals for $\phi^3$-theory), it's sufficient to apply our analysis on scalar integrals only. \cite{veltman79}


\section{Soft Singularity}

\begin{figure}[h]
\begin{center}
\includegraphics[width=6cm]{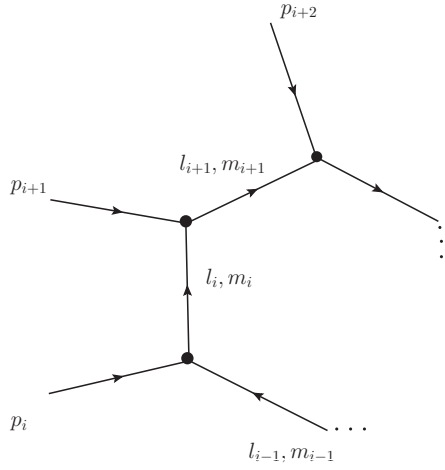}
\caption{ General scalar one-loop diagram with momentum assignment }
\label{fig:loopdiagram}
\end{center}
\end{figure}

These singularities appear as a result of any of the internal lines becoming soft, that is, it's momentum vanishes. We consider the $N$-point
scalar integral, Fig.~\ref{fig:loopdiagram}, in $D$-dimensions, given by

\begin{equation}
  I^{(N)}_D(p_i,m_i;i=0\rightarrow N-1) = \int d^Dl{ \frac{1}{ D_0 D_1 ..... D_{N-1}}},
\end{equation}

with following notations for simplification,
\begin{eqnarray}
 D_i &=& {l_i}^2 - {m_i}^2 ,\nonumber\\
 l_i &=& l+q_i \hspace{2.0 cm} \mbox{and} \nonumber\\
 q_i &=& p_0+p_1+....+p_i.
\end{eqnarray}
In our notation {\it $p_0 = 0$}, always but to start with, {\it $m_0 \neq 0$}. Now, we wish to derive conditions under which above integral may diverge as one of 
the internal momenta in the loop, say {\it $l_i$} becomes soft. We will take {\it $l_i = \epsilon $}, with the understanding that soft limit for {\it $l_i$} is reached as 
$\epsilon \rightarrow 0$. With {\it $l_i = \epsilon $}, in our notation, relevant denominators take following form 

\begin{eqnarray}
  D_{i-1} &=& p_i^2-2\epsilon \cdot p_i-m_{i-1}^2,\nonumber\\
  D_i &=& \epsilon^2-m_i^2 \hspace{2.0 cm} \mbox{and} \nonumber\\
D_{i+1} &=& p_{i+1}^2+2\epsilon \cdot p_{i+1}-m_{i+1}^2.
\end{eqnarray}
We have dropped $\epsilon^2$ against $\epsilon\cdot p_i$ and $\epsilon\cdot p_{i+1}$, assuming $\epsilon$ is not orthogonal to $p_i$ and $p_{i+1}$. The above denominators 
vanish under soft limit, if 
\begin{equation}
 m_i=0, p_i^2 = m_{i-1}^2 \hspace{0.2 cm}\mbox{and}\hspace{0.2 cm} p_{i+1}^2 = m_{i+1}^2.
\end{equation}

\begin{figure}[h]
\begin{center}
\includegraphics[width=5cm]{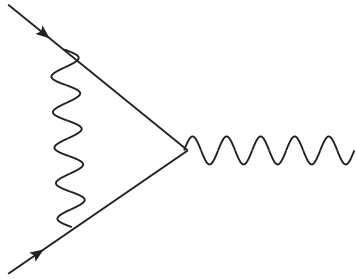}
\caption{ One loop correction to QED vertex }
\label{fig:vertex}
\end{center}
\end{figure}

Thus in soft limit scalar integral in Eq.(1) behaves as,

\begin{equation}
 I^{(N)}_D \sim \int d^D\epsilon {1\over \epsilon\cdot p_i\hspace{0.2 cm} \epsilon^2 \hspace{0.2 cm}\epsilon\cdot p_{i+1} } \sim \epsilon^{D-4},
\end{equation}
which diverges logarithmically in $D = 4$. In $m_i \rightarrow 0$ limit the divergence appears as ln $m_i$. Kinoshita ~\cite{Kinoshita62} called it 
$\lambda$-singularity. It is easy to check that no other denominator vanishes in soft limit of $l_i$, in general. Thus {\it the appearance
 of soft singularity in one loop diagrams is associated with the exchange of massless particles between two on-shell
particles}. The structure of soft singularity in Eq.(5) suggests that it can occur for $N\geq3$ point functions only. A text book example of soft singular
integral is the one loop vertex correction in QED with massive fermions, as shown in Fig.\ref{fig:vertex}. \cite{peskin95}


\section{Collinear Singularity}

At one loop, collinear singularity may appear when one of the internal momenta becomes collinear with a neighboring external leg. See Fig.1. We consider 

\begin{equation}
 l_i = x p_{i+1} + \epsilon_\perp,
\end{equation}
where $x\neq0,-1$ (since they correspond to softness of $l_i$ and $l_{i+1}$ respectively) and $\epsilon_\perp\cdot p_{i+1}=0$. The collinear limit is obtained 
as $\epsilon_\perp \rightarrow 0$. In this case relevant denominators are,

\begin{eqnarray}
D_i &=& x^2 p_{i+1}^2 + \epsilon_\perp^2 - m_i^2  \hspace{0.5cm} \mbox{and} \nonumber\\
D_{i+1} &=& (x+1)^2 p_{i+1}^2 + \epsilon_\perp^2 - m_{i+1}^2 . 
\end{eqnarray}

\begin{figure}[h]
\begin{center}
\includegraphics[width=6cm]{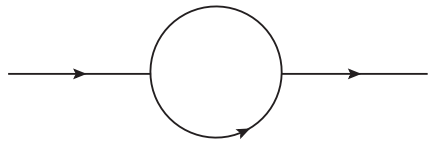}
\caption{ Two point function in $\phi^3$-theory }
\label{fig:bubble}
\end{center}
\end{figure}

The general conditions for these denominators to vanish are 

\begin{equation}
 p_{i+1}^2=0, \hspace{0.2cm} m_i = 0, \hspace{0.2cm} m_{i+1}=0,
\end{equation}
 that is, {\it one loop diagrams in which a massless external leg meets two massless internal lines may develop collinear divergence}.
In-fact, the first of the above three conditions is hidden in the assumption, $\epsilon_\perp\cdot p_{i+1}=0$. In $\epsilon_\perp \rightarrow 0$
limit, this equality implies

\begin{equation}
 \cos\theta \simeq {p_{i+1}^0 \over \vert {\bf p_{i+1}} \vert} > 1
\end{equation}
for massive $p_{i+1}$. Thus for collinear limit ($\theta = 0 $ ), $p_{i+1}$ must be massless. No other denominator vanishes for non-exceptional phase space points. 
The scalar integral (1), in this limit goes as

\begin{equation}
 I^{(N)}_D \sim \int d^D\epsilon_\perp {1\over \epsilon_\perp^2 \hspace{0.2 cm}\epsilon_\perp^2 } \sim \epsilon_\perp^{D-4},
\end{equation}
which, like soft singularity, is also logarithmically divergent. This singularity, sometimes referred as $m$-singularity can 
be regularized by setting $m_{i+1} = m_i$ and taking $m_i \rightarrow 0$ limit. It appears as ln $m_i$, as expected. We note that a two point function, Fig.\ref{fig:bubble}, can have IR singularity of collinear type only.  \\



\section{Overlapping Regions}

Now, since we have studied the structure of soft and collinear singularities of a one loop diagram , it is desirable to seek the 
possibility of their overlap. A soft piece, with all lines massless, contains two collinear pieces. Let this soft piece
be the part of a scalar $N$-point function in $D$-dimensions. We set $l_i=\eps $, to write the $N$-point scalar integral for massless internal and external lines, keeping only (in general) potentially 
divergent denominators as $\eps \rightarrow 0 $.

\begin{equation}
 I^N_D \simeq \int d^D\eps {\frac{1}{(\eps^2-2\eps\cdot p_i) \eps^2 (\eps^2+2\eps\cdot p_{i+1})}}.
\end{equation}

Instead of taking  $\eps \rightarrow 0 $ limit right away, which clearly corresponds to soft limit, we wish to break the above integral into soft and collinear regions. This
job is easily done in the light-cone co-ordinates as suggested in Ref.~\onlinecite{Sterman96}. In this co-ordinate, a 4-vector is written as $ v \equiv (v^+,v^-,{\bf{v}_\perp}) $ where,
$ v^\pm = \frac{1}{\sqrt{2}} (v^0\pm v^3)$ and ${\bf v_\perp} = (v^1,v^2)$ is a $2$-dimensional Euclidean vector. The dot product of two $4$-vectors then takes following form

\begin{equation}
 u\cdot v = u^+v^- + u^-v^+ - \bf{u_\perp}\cdot\bf{v_\perp}.
\end{equation}

This can easily be generalize in $D$-dimensions. In the CM frame of $p_i$ and $p_{i+1}$, using light-cone variables, we may write

\begin{eqnarray}
 p_i &\equiv& \sqrt{2} \omega (1,0, \bf{0_\perp}) \nonumber \\
 p_{i+1} &\equiv& \sqrt{2} \omega (0,1,\bf{0_\perp}).
\end{eqnarray}

Here $\bf{0_\perp}$ is ($D-2$) dimensional null vector in Euclidean space. Now, the integral in Eq.(11) reads
\begin{equation}
 I^N_D \simeq \int  {\frac{ d\eps^+ d\eps^- d^{D-2}{\bf{\eps}_\perp}}{(\eps^2-2\sqrt{2}\omega \eps^-) \eps^2 (\eps^2+2\sqrt{2}\omega \eps^+)}},
\end{equation}
with $\eps^2 = 2\eps^+\eps^- - {\bf \eps_\perp}^2$. Notice that we can make $\eps$ and therefore $l_i$, collinear to $p_i$ by setting $\eps^-$ and $\bf{\eps}_\perp$ equal to zero. 
To do it more systematically, we choose $\eps^- = \lambda {\eps_\perp}^2 $ where $\lambda \ne 0$ and take $\eps_\perp \rightarrow 0$ limit in Eq.(14), so that the integral becomes

\begin{equation}
  I^N_D \sim \int  {\frac{ d\eps^+ d\lambda {\bf{\eps_\perp}}^2 d^{D-2}{\bf{\eps}_\perp}}{{\bf \eps_\perp}^2 {\bf \eps_\perp}^2 \eps^+}} =
\int { d\lambda \frac{d\eps^+}{\eps^+} \frac{d^{D-2}\eps_\perp}{{\eps_\perp}^2} }
\end{equation}

Once again we obtain collinear singularity of log-type in $D=4$. Further, if we take $ \eps^+ \rightarrow 0 $, we make $l_i$ soft and we see the overlapping of 
soft and collinear singularities, which is also logarithmic in nature. Note that this singularity structure never gets worse for any other choice of $\eps^-$ made above.
It should be obvious that for non-exceptional phase space points there is no overlapping of two soft regions or two distant collinear regions in a one loop diagram. Thus
a one loop IR divergent integral is written as sum of terms, each containing two large-log factors, at most.
In dimensional regularization ($D=4-2\eps_{IR}, \eps_{IR} \rightarrow 0^-$), IR singular terms at one loop appear as coefficients of $ 1\over \epsilon_{IR}$ (soft and/or collinear case)
and $1\over\epsilon_{IR}^2$ (overlapping case). \cite{Ellis08} In Fig.\ref{fig:vertex}, if we allow fermion lines to be massless, the one loop correction to 
QED vertex exhibits the full structure of IR divergence at one loop.~\cite{Sterman96} \\  

\section{Conclusion}

We have shown many known features of IR divergence of a one loop diagram, using naive power counting in loop momentum. We have seen that for a one loop diagram to have IR 
singularity at least one internal line must be massless. Diagrams with all external legs 
massive are IR finite for all internal lines massless. Tensor integrals at most retain the IR singular structure of scalar integrals. Since maximum three denominators of a general $N$-point
scalar integral vanish at a time, in infrared regions, one should expect the possibility of expressing IR singular terms of any one loop diagram ($N>3$), in terms of those of appropriate three point functions.
In Ref.~\onlinecite{Dittmaier03}, this expectation is achieved for the most general one loop integrals. The above analysis also tells us that any one loop diagram is IR finite in $D>4$ dimensions. 
This is useful in calculation of certain one loop integrals in $4$-dimensions in terms of those in $6$-dimensions which certainly contributes to finite part of the integral and can be evaluated numerically. \cite{Bern93}


\end{document}